\documentclass[11pt,onecolumn,showpacs]{revtex4}
\usepackage{graphicx, color}
\usepackage{verbatim}

\newcommand{\tabrule}{\rule[-0.2em]{0em}{1.2em}}

\def\ind#1{{_{\mathrm{#1}}}}

\begin{document}

\title{Theoretical proposal for the dynamical control of the nonlinear optical response frequency}
\author{L. Chotorlishvili$^1$, A. Sukhov$^1$, S. Wimberger$^2$, and J. Berakdar$^1$}
\affiliation{$^1$Institut f\"{u}r Physik, Martin-Luther Universit\"{a}t Halle-Wittenberg, Heinrich-Damerow-Straße 4, 06120 Halle, Germany \\
$^2$Institut f\"{u}r Theoretische Physik and Center for Quantum Dynamics, Universit\"{a}t Heidelberg, 69120 Heidelberg, Germany}
\date{\today}

\begin{abstract}
We propose theoretically a method for  the control of the frequency of the nonlinear optical response of a model 
nonlinear medium driven by an electric field
  and suggest an 
 experimental realization on the basis of  isotropic $LiF$,  or $NaCl$ crystals  and pulsed $Nd$ glass laser. The theoretical background of the proposed optical control is  the nonlinear resonance and the resonance overlapping which is typical for nonlinear systems. It is shown that by a proper choice of the parameters of the applied pulses a diffusive growth of the oscillation amplitude can be achieved resulting in a controlled switching of the system's frequency. The maximum frequency of the nonlinear response is identified to be proportional to the intensity of the applied noise. An experimental suggestion for particular materials is proposed.
\end{abstract}

\pacs{...}

\maketitle

\section{Introduction}
With the accumulated  volume of information that has to be transmitted, assorted and stored swiftly and  efficiently  there is
growing demand to utilize for these purposes   methods and techniques  from  nonlinear \cite{Shen02,Yari76,Bloe65,ScWi86} and fiber optics \cite{Agraw01,EiAn05,Argy05,Holz11,KuBr11,AbNa09}. \\
For an optical system that is slightly off its  equilibrium state, the model of a linear oscillator is a well-established picture that yields a good qualitative and quantitative agreement between theory and experiment. In particular, for different types of optical media, optical characteristics such as the electric susceptibility, the polarization and  the transmission-refraction coefficients were  predicted  correctly by the linear oscillator model \cite{Shen02,Yari76,Bloe65,ScWi86}. However, systems that are far off the equilibrium possess usually  physical features that cannot be explained using linear models.  For example, the phenomenon of the self-induced transparency and the generation of higher harmonics were successfully explained by a nonlinear oscillator model \cite{EiAn05}.  The problem of the light-induced critical behavior in nonlinear systems is a classical problem and was addressed in many studies \cite{FlTa80,KiYa84,RiBo85}. An interesting phenomenon observed is the polarization related optical bistability that occurs due to the dependence of  the  transmission and  the  absorption coefficients on the oscillation amplitude that may well be chaotic for a nonlinear system. The polarization dynamics is usually defined by the Kitano map \cite{KiYa84} and undergoes a bifurcation transition with the increase of light intensity. However, the problem of the optical transistor we are going to address in the present study is not connected with the polarization dynamics, but with the controlled switching of the frequency. The physical mechanism for the functioning of the proposed optical transistor is the phenomenon of nonlinear resonance and resonance overlapping. This phenomenon is typical for externally driven nonlinear systems and can be utilized for the diffusive control of the frequency of system. For  nonlinear optical media we may expect a response with the frequency being several times higher than the frequency of the applied external driving field. The question that arises naturally is whether it is possible to tune and control the outgoing frequency. Here, we develop an efficient frequency switching scheme which can be easily realized  experimentally. We focus on reaching the maximum desired outgoing frequency that defines the maximum transmission bandwidth for the optical transistor. We also address an even more important issue - the freezing of the target frequency fixed  which naturally required
 for the stability of the device. \\
The paper is organized as follows: In the second section we specify the problem for the case of a single oscillator model and evaluate the maximum possible response frequency which can be obtained from the optical media using a monochromatic external driving field. We also consider the scheme based on the polychromatic pumping technique. We utilize the method of nonlinear resonance in order to derive the corresponding Fokker-Planck equation that describes the evolution of the system frequency when the oscillator dynamics is chaotic in time.  Additionally, we study the consequences of both dissipation and thermal noise. In the third section we generalize the model for the case of interacting oscillators paying a special attention to the collective dynamics and   the correlation effects between different oscillators.

\section{Single oscillator model}

\subsection{Model}

Generally the oscillator model views the active medium as consisting of   a set of classical anharmonic
oscillators in a unit volume. Each oscillator stands  physically for an electron
bound to a core or an infrared-active molecular vibration.
The total energy of the  electron in the presence of a driving electric  field reads \cite{Shen02}
\begin{equation}
\begin{array}{l}
 \displaystyle H=H_0 \left( {x,p} \right) + H\ind{NL} \left( x \right) + \varepsilon V\left( {x,t} \right),  \\
 \displaystyle H_0=\frac{{p^2 }}
{{2m}} + \frac{{\omega _0^2 mx^2 }}
{2},  \\
 \displaystyle H\ind{NL}=- \alpha x^3  - \beta x^4 .
\end{array}
\label{eq_1}
\end{equation}
Here $H_0 \left( {x,p} \right)$ is the energy of the harmonic oscillator, $H\ind{NL}$ is a nonlinear correction of energy, and $\varepsilon V(x,t)$ is the interaction with the electromagnetic field. In particular we will consider two possible cases for the external driving: a) an applied external monochromatic electric field $\varepsilon V\left( {x,t} \right) = xV_0 (t),\,\,\,\,\,V_0 (t) = ef\cos (\Omega t),$ with the amplitude $f$ and frequency $\Omega$, and b) an infinite series of applied delta pulses $V\left( {x,t} \right) = xV_0 \sum_{n =  - \infty }^\infty  {\delta (t - nT)}$, $V_0  = ef$, with the amplitude $f$ and the interval between the pulses being equal to $T$. In Eq. (\ref{eq_1}) $e$ and $m$ are the charge and the mass of the electron, $\alpha$ and $\beta$ are the nonlinearity coefficients, $x$ and $p$  are the coordinate and the momentum of the electron, and  $\omega_0$ is the eigenfrequency of harmonic oscillations. In what follows we assume that the eigenfrequency $\omega_0$  and the driving frequency of the monochromatic field $\Omega$  are optical frequencies, while the period between the pulses is chosen to be of the order of a picosecond. Therefore, the following condition is met $\frac{{2\pi }} {{\omega _0 }},\,\,\frac{{2\pi }}{\Omega } \ll T$. In what follows, we set time interval between pulses $T$  as the unit of the time and therefore $T=1$. These parameters are typical for the  Nd-gas laser \cite{ScWi86} operating in the mode-locking regime. The modeling of the driven physical system with a nonlinear oscillator model seems a great simplification \cite{Shen02}. However, as mentioned in the introduction it has been  established as a very efficient and to a large extent reliable approach in nonlinear optics \cite{UgMc11}.

\subsection{Monochromatic driving}

In the case of a monochromatic driving, we add to the equation of motion (\ref{eq_1}) a phenomenological decay term $\gamma \dot{x}$ responsible for thermal loses in the system (\ref{eq_1})
\begin{equation}
\ddot x =  - \omega _0^2 x -\frac{3\alpha}{m}x^2- \frac{{4\beta }}
{m}x^3  - \frac{\gamma }
{m}\dot x + \frac{{V_0 }}
{m}\cos \left( {\Omega t} \right).
\label{eq_2}
\end{equation}
In the case of a weak nonlinearity corresponding to the state that deviates slightly  from the equilibrium, i.e. $\left( {x(t)} \right)_{\max }  < \frac{{\omega _0 }}{2}\sqrt {\frac{m}{\beta }} $, equation (2) can be solved analytically using canonical perturbation theory \cite{LaLi76,LL92}. Consequently, for the amplitude frequency characteristics one obtains the following expressions

$$
\omega  = \omega _0  + \bigg(\frac{{3\beta }}
{{8\omega _0 m}}-\frac{5\alpha^{2}}{12m\omega_{0}^{3}}\bigg)A_{\max }^2 .
$$

\begin{figure}
\centering \includegraphics[scale=.55,angle=1]{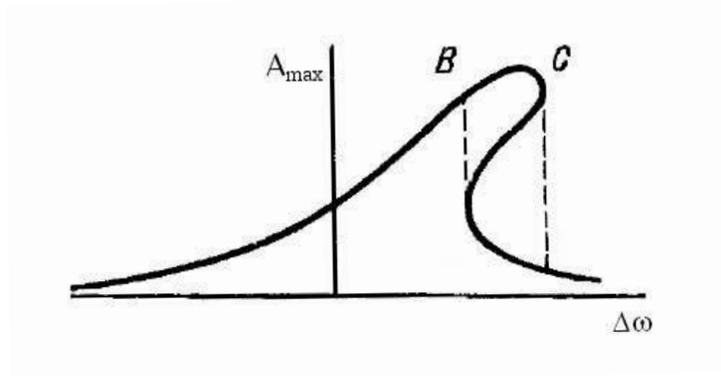}
\caption{Schematically shown dependence of the oscillation amplitude $A_{\max}$ \cite{LaLi76} on the  detuning between the frequency of the external field and the frequency of the linear oscillations $\Delta \omega=\Omega-\omega_0$.  The maximum value of the amplitude belongs to the unstable domain $BC$ of the hysteresis loop. }
\label{fig_1}
\end{figure}

In case of weak nonlinearity maximal values of the frequency shift corresponds to the case $\alpha =0$:
\begin{equation}
\omega _{\max }  = \omega _0  + \frac{{3\beta }}
{{8\omega _0 m}}A_{\max }^2 ,
\label{eq_3}
\end{equation}
where $A_{\max }^2$ is the maximum square of the amplitude which may be achieved during the oscillations for the case $\alpha=0$
\begin{equation}
A_{\max }^2  = \frac{{V_0^2 }}
{{4\omega _0^2 \gamma ^2 }}.
\label{eq_4}
\end{equation}
Since the maximum values of the oscillation amplitude is limited by the condition (\ref{eq_4}), the maximum nonlinear frequency shift $3\beta/(8\omega_0 m)A^2_{\max}$  and, consequently, the maximum frequency $\omega_{\max}$ that can be achieved using the monochromatic driving external field is limited too.  Another drawback of the monochromatic driving is the fact that the maximum frequency $\omega_{\max}$ belongs to the unstable domain of the amplitude frequency characteristics (cf. Fig. \ref{fig_1}). Hence, even a small fluctuation of the driving frequency $\Delta\omega=\Omega-\omega_0$ leads to a drop of the maximal frequency $\omega_{\max}$ down to the linear frequency $\omega_0$.
\subsection{Dynamical stochasticity and diffusive control of the frequency}
Looking for a more efficient frequency switching scheme for the optical transistor,  we consider an infinite series of delta-function pulses (kicks) applied to the system, i.e. $V\left( {x,t} \right) = xV_0 T\sum_{n =  - \infty }^{\infty}  {\delta (t - nT)}$, where $V_0  = ef$. First, we neglect dissipation in the system. Transforming to the canonical variables $\left(I,\theta \right)$ of the harmonic oscillator (with frequency $\omega_0$), i.e., $x = \sqrt {\frac{{2I}}{{m\omega _0 }}} \cos \theta$ and $p = \sqrt {2I\omega _0 m} \sin \theta $, we deduce from eq. (1)
that after averaging over the fast phases \cite{UgMc11}, the relations apply
\begin{equation}
\begin{array}{l}
\displaystyle H=H_0 (I) + V(I,\theta ,t),  \\
  \displaystyle H_0(I)= I\omega _0  + 3\pi \beta \left( {\frac{I}{{m\omega _0 }}} \right)^2, \\
  \displaystyle V(I,\theta ,t)= V(I,\theta )T\sum\limits_{n =  - \infty }^\infty  {\delta (t - Tn)}, \\
   \displaystyle \mathrm{where} \,\, V(I,\theta ) = V_0 (I)\cos \theta \,\, \mathrm{and} \,\, V = V_0 (I) \approx V_0 \sqrt {\frac{{I(0)}}{{m\omega _0 }}}, \\
  \displaystyle \omega (I) = \omega _0  + \frac{{6\pi \beta }}{{m^2 \omega _0^2 }}I.
\end{array}
\label{eq_5}
\end{equation}
From the system of eqs. (\ref{eq_5}) we see that the frequency of the system depends on the action $\omega (I)$. Consequently, when controlling $I$, we can control the frequency of the system. The Hamilton equations of motion corresponding to the energy given by eqs. (\ref{eq_5}) read
\begin{equation}
\begin{array}{l}
\displaystyle \dot I =  - \varepsilon \frac{{\partial V\left( {I,\theta ,t} \right)}}{{\partial \theta }},  \\
\displaystyle  \dot \theta  = \omega (I) + \varepsilon \frac{{\partial V\left( {I,\theta ,t} \right)}}{{\partial I}}.  \\
\end{array}
\label{eq_6}
\end{equation}
The benefit of using the short kicks is the simplicity of the corresponding equations of motion (\ref{eq_6}). Since the motion is governed by either kicks or by the free evolution inbetween two kicks, the equations of motion (\ref{eq_6}) can be integrated straightforwardly over one temporal period by splitting the evolution operator in two parts $\hat T = \hat T_R \hat T_\delta $.  Here
\begin{equation}
\hat T_R \left( {I,\theta } \right) = \left( {I,\theta  + \omega \left( I \right)T} \right)
\label{eq_7}
\end{equation}
is the evolution operator responsible for the free rotations between the pulses, while the operator $\hat{T}_{\delta}$ describes the influence of the pulses
\begin{equation}
\hat T_\delta  \left( {I,\theta } \right) = \left( {I - \varepsilon T\frac{{\partial V(I,\theta )}}{{\partial \theta }},\,\,\,\,\theta  + \varepsilon T\frac{{\partial V(I,\theta )}}{{\partial I}}} \right).
\label{eq_8}
\end{equation}
Taking into account eqs. (\ref{eq_7}) and (\ref{eq_8}) and assuming that ${{6\pi \beta }}/({m^2 \omega _0^2 }) \gg {{V_0 }}/({{2\sqrt {m\omega _0 } }})$, the solution of the equations of motion (\ref{eq_6}) can be found in the form of the standard map \cite{Zasl07}
\begin{equation}
\begin{array}{l}
\displaystyle I_{n + 1}  = I_n  + K\sin \theta , \\
\displaystyle \theta _{n + 1}  = \omega _0 T + \theta _n  + I_{n + 1},
\end{array}
\label{eq_9}
\end{equation}
where $K \approx V\frac{{6\pi \beta }}{{m^2 \omega _0^2 }}T^2 $.\\

\begin{figure}
\centering \includegraphics[scale=.45]{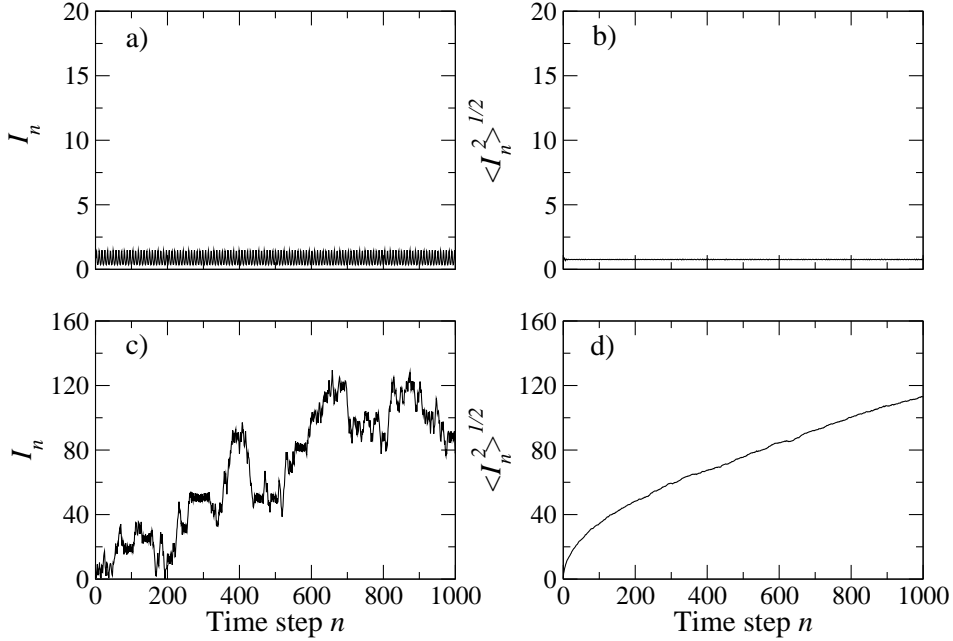}
\caption{$I_n$ dynamics in the dimensionless units $\frac{\sqrt{<I^{2}(t)>}}{I_{0}}=\sqrt{1+D't}=\sqrt{1+t/T}$ on the scale $n\in [0:1000]$ for $\gamma=0$. a)
    and b) correspond to the stochasticity coefficient $K=0.5$; c) and d)
    describe the dynamics for $K=5.0$. $<I^2_n>^{1/2}$ (b, d) is defined as the
    ensemble average over $1000$ realizations of the initial random values from the intervals $I_{n=0}
    \in (0:1)$ and $\theta_{n=0} \in (0:2\pi)$.}
\label{fig_2}
\end{figure}

Despite its relative simplicity, the standard map displays a complicated and rich behavior \cite{Zasl07,LL92,BeKa97,SW2011}. For large $K>1$ the phase space of the system is predominantly chaotic and is an invariant Kolmogorov-Arnold-Moser (KAM) torus \cite{Zasl07,BeKa97}.
In the chaotic sea that covers most of the phase space, the dynamics of the action variable $I$ is expected to be diffusive in time. Taking into account the following values of parameters that corresponds to the Iceland spark (Iceland crystal) and the isotropic crystals $LiF$, $NaCl$ and $Nd$ glass lasers \cite{Yari76,Bloe65} \cite{UgMc11} $T=10^{-12}$~[s], $\omega_0=10^{15}$~[s$^{-1}$], $m=10^{-31}$~[kg], $e=10^{-19}$~[C], $\beta=4\cdot 10^{11}$~[kg/(m$^2$s$^2$], $x_{\max } (0) \sim 10^{ - 8}$~[m], we see that the condition $K>4$ imposes the following restriction on the driving field amplitude
\begin{equation}
f > \frac{1}
{e}\frac{{4m^2 \omega _0^2 }}
{{6\pi \beta T^2 x_{\max } (0)}} \approx 10^7 \mathrm{~[V/m]}.
\label{eq_13}
\end{equation}
With increasing $K$ the principal resonance island gradually becomes smaller and disappears for about $K \gtrsim 5$ \cite{LL92}.
 When the condition $K>1$
holds, then the dynamics of the action variable is chaotic and is described by the following Fokker-Planck equation \cite{Zasl07}
\begin{equation}
\frac{{\partial F(I)}}{{\partial t}} =  - \frac{\partial }{{\partial I}}\left( {AF(I)} \right) + \frac{{\partial ^2 }}{{\partial I^2 }}\left( {DF(I)} \right),
\label{eq_18}
\end{equation}
where $A = \left\langle {\Delta I} \right\rangle$, $D = \left\langle {\left( {\Delta I} \right)^2 } \right\rangle$, $\Delta I = I_{n + 1}  - I_n $. Taking into account eqs. (\ref{eq_9}) and (\ref{eq_18}) we find
\begin{equation}
A = 0, \,\,\, D = T\left\langle {\left( {V\sin \theta } \right)^2 } \right\rangle  = T\frac{{V^2 }}{2}.
\label{eq_19}
\end{equation}
The Fokker-Planck equation can be simplified as
\begin{equation}
\frac{{\partial F}}{{\partial t}} = \frac{1}{2}D\frac{{\partial ^2 F}}{{\partial I^2 }}
\label{eq_20}
\end{equation}
and finally we obtain
\begin{equation}
\sqrt {\left\langle {I^2 } \right\rangle }  = \sqrt {I_0^2  + Dt}.
\label{eq_21}
\end{equation}
Using the connection between the action variable and the nonlinear frequency (\ref{eq_5}) and (\ref{eq_21}) we derive an explicit expression for the time dependence of the nonlinear frequency
\begin{equation}
\left\langle {\omega (I)} \right\rangle  = \omega \left( {\sqrt {\left\langle {I^2 } \right\rangle } } \right) = \omega _0  + \frac{K}{{VT^2 }}\sqrt {I_0^2  + Dt}  = \omega _0  + \frac{{6\pi \beta }}{{m^2 \omega _0^2 }}I_0 \sqrt {1 + D't} .
\label{eq_22}
\end{equation}
Here $D' = \frac{{TV^2 }}{{2I_0^2 }}$, $I_0  = \omega _0 mx_{max}^{2} \approx 10^{ - 32}$~[Js]. Taking the values of the parameters from Ref. \cite{UgMc11}, i.e. $T=10^{-12}$~[s], $\omega_0=10^{15}$~[s$-1$], $m=10^{-31}$~[kg], $e=10^{-19}$~[C], $\beta=4\cdot 10^{11}$~[kg/(m$^2$s$^2$)], $x_{\max } (0) \sim 10^{ - 8}$~[m], we arrive at the following estimate $\frac{{6\pi \beta }}{{m^2 \omega _0^2 }}I_0  \approx 24\pi \omega _0$, $D'\approx 10^{12}$~[s$^{-1}]=1/T$.\\
The meaning of the expression (\ref{eq_22}) is the following. Imagine, that the pulses of a light beam are incident on a nonlinear optical medium which is modeled as an ensemble of nonlinear non-interacting oscillators. The light heats up the oscillator motion and converts the regular small oscillations with the harmonic frequency $\omega_0$ into the irregular oscillations with the nonlinear frequency $\omega (I)$. An averaging using the Fokker-Planck equation is equivalent to the averaging over an ensemble of non-interacting chaotic oscillators. As a result, the averaged quantity $<\omega (I)>$ exactly matches the frequency of the resulting outgoing signal that could be expected from optical media in an experiment for real physical materials. For example, for Iceland spark and isotropic crystals such as $LiF$ nad $NaCl$, the model of a driven nonlinear oscillator gives the theoretical estimates that are in a good qualitative and quantitative agreement with the experimentally observed physical properties \cite{Bloe65,ScWi86}. Suppose, that our aim is to achieve a signal from an optical medium with the frequency that is a multiple of the initial harmonic frequency $\omega_0$, $N\omega_0$, where $N$ is an integer number.  According to Eq. (\ref{eq_22}), as a consequence of the stochastic heating, the nonlinear optical medium responds at  $\omega=N\omega_0$ after a time defined via the following equation
\begin{equation}
t_c  = \frac{1}
{D'}\left[ \frac{T^{4}}{I_0^2}{\left( {\frac{{V }}{K}} \right)^2 \omega _0^2 \left( {N - 1} \right)^2  - 1 } \right].
\label{eq_23}
\end{equation}
Eq. (\ref{eq_23}) corresponds to the normal diffusion case. In the case of a superdiffusion it modifies to
\begin{equation}
t_c  = \frac{1}{{D'}}\left[ \frac{T^{4}}{I_0^2}{\left( {\frac{{V}}{K}} \right)^2 \omega _0^2 \left( {N - 1} \right)^2  - 1 }\right]^{1/\mu } ,\,\,\,1 < \mu  < 2\,
\label{eq_24}
\end{equation}
The analytical result given by eq. (\ref{eq_23}) is in a good agreement with the exact numerical calculations for the map given by eq. (\ref{eq_9}) (cf. Fig. \ref{fig_2}).  Inserting standard values of the parameters (listed below the Eq. (22)) in the Eq. (23) provides a very simple estimation for the frequency switching time: $t_{c}=10^{4}T\left(\frac{N-1}{K}\right)^{2},~~~~<\omega(t)>=\omega_{0}+\omega_{0}10^{-2}K\sqrt{1+D't},~~~~   \frac{\sqrt{<I^{2}(t)>}}{I_{0}}=\sqrt{1+D't},~~~D'=1/T$.  Now we immediately can estimate time $t_c$ (number of kicks, since unit of time is the interval between pulses) for the switching of initial frequency $\omega_{0}$ to the final frequency $\omega=N\omega_{0}$, for a given stochastic parameter $K$.

Indeed, the numerical calculations confirm the diffusive increase of the action variable for this case.\\
Note that in eqs. (\ref{eq_23}) and (\ref{eq_24}) we do not set any limitations on $N$. This means, that one can reach a final frequency as large as one wishes in principle. This is a principle benefit of the applied pulses, since in the case of a monochromatic pumping we have a maximum frequency limit (eq. (\ref{eq_3})). Nevertheless, the disadvantage of eqs. (\ref{eq_23}) and (\ref{eq_24}) is the frequency stabilization problem. Due to the diffusion (by the nonlinear nature of the problem) the frequency will increase even after reaching the desired value $\omega=N\omega_0$. However, as will be shown below, for more realistic models which include the decaying and the thermal effects we can circumvent this problem.
\subsection{Dissipative map and frequency freeze}
Obviously, a realistic model should include not only the anharmonicity of the oscillations but also the dissipation that may happen in optical media due to thermal losses. We therefore generalize our model (eq. (\ref{eq_5})) by adding a phenomenological damping term $-\gamma \dot{x}$. Following the standard procedure given by eqs. (\ref{eq_7}) and (\ref{eq_8}) and splitting the evolution operator in two parts $\hat T = \hat T_R \hat T_\delta$, it can be shown that the dissipation term modifies the standard map in the following way \cite{Zasl07}
\begin{equation}
\begin{array}{l}
\displaystyle I_{n + 1}  = e^{ - \gamma } I_n  + K(\gamma )\sin \theta _n,  \\
\displaystyle  \theta _{n + 1}  = \theta _n  + I_{n + 1},  \\
\displaystyle  K(\gamma ) = \left( {\frac{{1 - e^{ - \gamma } }}{\gamma }} \right)K.
\end{array}
\label{eq_25}
\end{equation}
Here $I=(I-I_0)/I_0$ and $\gamma=\gamma T$ are the rescaled dimensionless action variable and the rescaled decay constant, respectively.The Fokker-Planck equation has the form of eq. (\ref{eq_18}), however, the coefficients given by eq. (\ref{eq_19}) take on a different form
\begin{equation}
\displaystyle A = \left( {1 - e^{ - \gamma } } \right)I,\,\,\,D = \left( {1 - e^{ - \gamma } } \right)^2 I^2  + \frac{1}{2}K^2 e^{ - 2\gamma }
\label{eq_26}
\end{equation}
Taking into account eq. (\ref{eq_26}), the Fokker-Planck equation reads
\begin{equation}
\displaystyle \frac{{\partial F(I)}}{{\partial t}} = \left( {1 - e^{ - \gamma } } \right)\left(\frac{\partial }{{\partial I}}\left( {IF(I)} \right) + \frac{1}{2}\frac{{\partial ^2 }}{{\partial I^2 }}\left( {I^2 F(I)} \right)\right) + \frac{1}{4}K^2 e^{ - 2\gamma } \frac{{\partial ^2 F(I)}}{{\partial I^2 }}.
\label{eq_27}
\end{equation}

\begin{figure}
\centering \includegraphics[scale=.45]{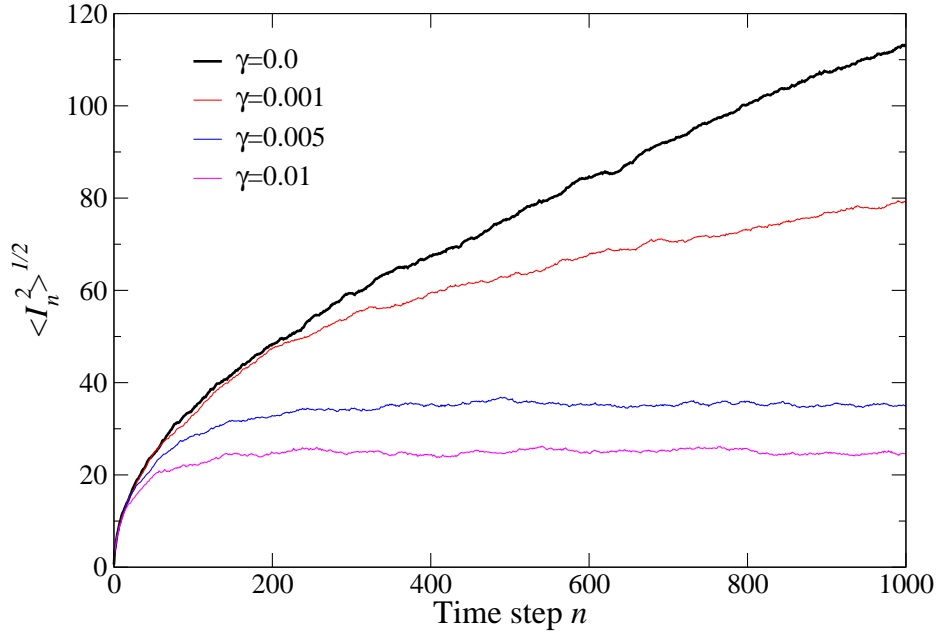}
\caption{$I_n$ dynamics on the scale $n\in [0:1000]$ for $K=5.0$ and various $\gamma$. $\sqrt{<I^2_n>}$ is defined as the
    ensemble average over $1000$ realizations of the initial random values from the intervals $I_0
    \in (0:1)$ and $\theta_0 \in (0:2\pi)$.
    For finite $\gamma$ the theory predicts $\displaystyle
    \sqrt{<I^2_n>}=\frac{K}{2\sqrt{\gamma}}|_{t\rightarrow \infty}$ (eq. \ref{eq_29}). For
    $\gamma=0.005$ and $\gamma=0.01$ the analytical expression yields
    $\sqrt{<I^2_n>}\simeq 25$ and $\sqrt{<I^2_n>} \simeq 35$, respectively. The
    numerical calculations illustrated by the figure for $\gamma=0.005$ and
    $\gamma=0.01$ visibly saturate to those values.
    The time needed to approach the saturation value scales as $\approx
    1/\gamma$.
}
\label{fig_4}
\end{figure}

Eq. (\ref{eq_27}) is quite involved and therefore finding of  exact analytic solution for distribution function is
complicated problem. However for evaluation of the second moment of random variable $<I^{2}>=\int_0^\infty dI\cdot I^{2}F\big(I,t\big)$ we do not need explicit solution for distribution function. Multiplying both sides of  eq. (\ref{eq_27}) on $I^{2}$, integrating over  $I$ and taking into account
$\int_0^\infty \frac{{\partial F(I)}}{{\partial t}}I^{2}dI =\frac{\partial}{\partial t}<I^{2}>$ we deduce:
\begin{equation}
\displaystyle \left\langle {I^2 } \right\rangle  = I_0^2 \exp \left( { - 2\gamma t} \right) + \frac{1}
{{4\gamma }}K^2 \left[ {1 - \exp \left( { - 2\gamma t} \right)} \right],
\label{eq_28}
\end{equation}
with the asymptotic limit
\begin{equation}
\begin{array}{l}
\displaystyle  \sqrt {\left\langle {I^2 } \right\rangle }  = \sqrt {\left\langle {\frac{{I - I_0 }}{{I_0 }}} \right\rangle ^2 }  = \frac{K}{{2\sqrt \gamma  }}\,\,\,\,\,\,t \to \infty ,\,\,  \\
\displaystyle  \sqrt {\left\langle {\Delta I^2 } \right\rangle }  = \frac{{I_0 K}}{{2\sqrt \gamma}}.
\end{array}
\label{eq_29}
\end{equation}
Consequently, for the nonlinear frequency we obtain
\begin{equation}
\displaystyle \omega _{\max } \left( {t \to \infty } \right) = \omega _0  + \frac{{6\pi \beta }}{{m^2 \omega _0^2 }}I_0 \sqrt {1 + \frac{{K^2 }}{{4\gamma }}}  \approx \omega _0  + 24\pi \omega _0 \sqrt {1 + \frac{{K^2 }}{{4\gamma }}}.
\label{eq_30}
\end{equation}
From eq. (\ref{eq_30}) we infer that in the case of dissipative systems, the maximum frequency $\omega _{\max } \left( {t \to \infty } \right)$ has saturated values  that can be controlled by tuning  the chaos parameter $K$. The analytical results are in a very good quantitative agreement with the results of exact numerical calculations (cf. Fig. \ref{fig_4}).
\subsection{Noise-induced enhancement}
To finally close the consideration of the single oscillator problem, we discuss the effect of the controlled noise applied to the system. We consider a modification of the interaction, when the fixed kick amplitude is replaced by a step-dependent random amplitude and instead of eqs. (\ref{eq_25}) we obtain now:
\begin{equation}
\begin{array}{l}
\displaystyle I_{n + 1} (m) = e^{ - \gamma } I_n  + K(\gamma ,\,\,\xi _n )\sin \theta _n, \\
\displaystyle \theta _{n + 1} (m) = \theta _n  + I_{n + 1}, \\
\displaystyle K(\gamma ,\,\,\xi _n ) = \left( {\frac{{1 - e^{ - \gamma } }}{\gamma }} \right)\left( {K + \xi _n } \right),
\end{array}
\label{eq_31}
\end{equation}
\textcolor{black}{where $\xi_n$ is a Gaussian white noise $\xi(t),~~\langle\xi(t)\rangle=0$, with the  noise strength $q$  $\left<\xi(t+\Delta t)\xi(t)\right>=q \delta(\Delta t)$ which is implemented into the numerical procedure via standard methods, see e.g. \cite{Kamp07}}.

\begin{figure}
    \centering
    \includegraphics[width=.74\textwidth]{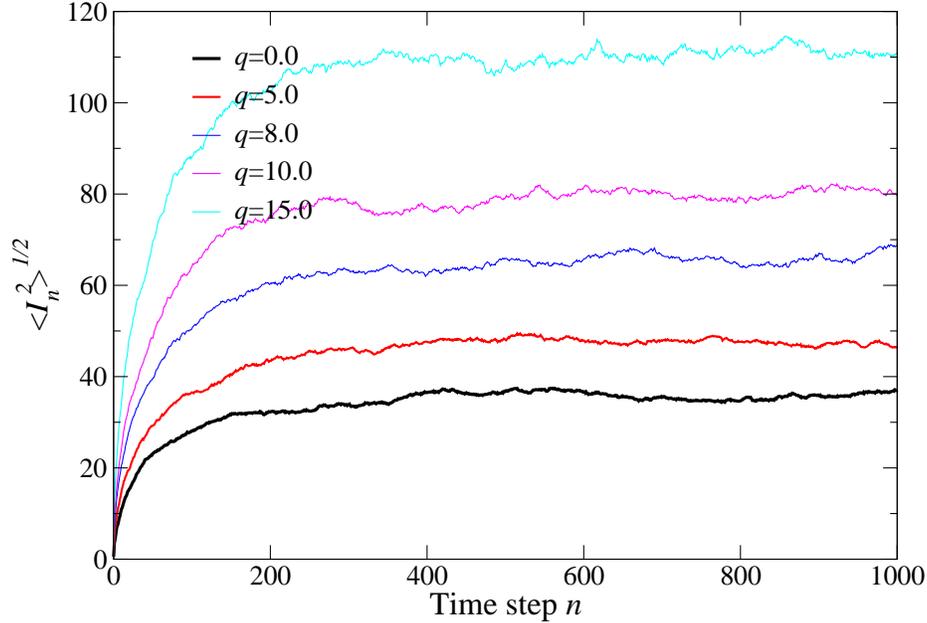}
    \caption{$I_n$ dynamics on the scale $n\in [0:1000]$ for fixed $K=5.0$ and
    $\gamma=0.005$. $\sqrt{<I^2_n>}$ is defined as the
    ensemble average over $1000$ realizations of the initial random values from the intervals $I_0
    \in (0:1)$ and $\theta_0 \in (0:2\pi)$. The curve with $q=0.0$ corresponds to that shown in Fig. \ref{fig_4}
     for the same $\gamma$. $q$ is the noise strength, i.e. $\left<\xi(t+\Delta t)\xi(t)\right>=q \delta(\Delta t)$.}
    \label{fig_5}
\end{figure}

\begin{figure}
    \centering
    \includegraphics[width=.74\textwidth]{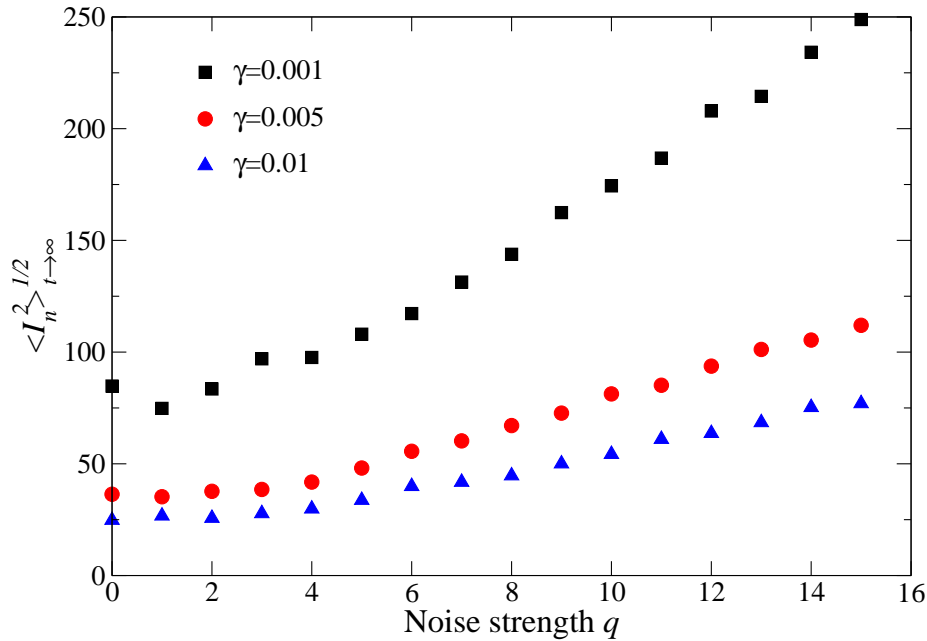}
    \caption{$\displaystyle \sqrt{<I^2_n>}|_{t\rightarrow
    \infty}(q)$-dependence obtained as a linear fit for regions where
    $\sqrt{<I^2_n>}$ saturates. The approximate values of $\alpha$ for the corresponding increasing damping parameter are found using eq. (\ref{eq_32}) and are $\alpha_{0} (\gamma=0.001)\approx 19$, $\alpha_{0} (\gamma=0.005)\approx 9$ and $\alpha_{0} (\gamma=0.01)\approx 7$.}
    \label{fig_6}
    \end{figure}

It is well-known that noise is sometimes profitable and can indeed enhance signals in many cases \cite{GaHa98}. 
We intend to explore the consequences of the applied noise on the transistor frequency. With this purpose we numerically integrate the recurrence relations (\ref{eq_31}) for different strengths of the applied noise. The results of the numerical calculations are presented in Figs. \ref{fig_5} and \ref{fig_6}.\\
From Fig. \ref{fig_6} we see that the saturation time is noise-independent while the saturated values of action increase with the noise strength. \textcolor{black}{An increase of the frequency upon enhancing  the noise is also obtained in other physical phenomena. Particularly, in experiments using the ferromagnetic resonance (FMR) technique \cite{AnLi05} the resonance magnetic field, which is associated with the resonance frequency, grows with the increasing noise level. Those experimental findings are well reproduced by numerical simulations including thermal noise \cite{Usad06,SuUs08,Denisov}.}\\
In addition, the effective stochastic coefficient $K$ turns out to be proportional to the noise strength $q$. This is obvious from the relation (\ref{eq_29}) $\sqrt {\left\langle {I^2 }\right\rangle }  = \frac{K}{{2\sqrt \gamma }}$ and the numerical data presented in Fig. \ref{fig_6}. We obtain a linear function of the noise $K(q) = K + \alpha_{0} q$, where the constant $\alpha_{0}$ can be extracted from Fig. \ref{fig_6}. Consequently, for the maximum frequency of the optical transistor we have the following estimation
\begin{equation}
\begin{array}{l}
\displaystyle \omega _{\max } \left( {t \to \infty } \right) \approx \omega _0  + 24\pi \omega _0 \sqrt {1 + \frac{{\left( {K + \alpha_{0} q} \right)^2 }}{{4\gamma }}} ,\\
\displaystyle K = ef\frac{{x_m 6\pi \beta }}{{m^2 \omega _0^2 }}T^2.
\end{array}
\label{eq_32}
\end{equation}
Now we have all ingredients concerning the optical transistor factor. It looks relatively simple and should easily be accessible in the experiment
\begin{equation}
\displaystyle N = \frac{{\omega _{\max } \left( {t \to \infty } \right)}}{{\omega _0 }} = 1 + 24\pi \sqrt {1 + \frac{{\left( {K + \alpha_{0} q} \right)^2 }}{{4\gamma }}}.
\label{eq_33}
\end{equation}
The only control parameters are the parameters of the applied pulses: the time interval between the pulses $T$, the amplitude of the pulses $f$ and the intensity of the applied noise.

Before generalize our approach for the ensemble of oscillators, we will consider role of fluctuations in order to have clear understanding how precisely switching procedure can be performed. Note that nonlinear frequency is a linear function of action $\omega\big(I\big)$  (5). Along with the diffusive growth of mean value of action  $\big<I\big>$ actual values of action  fluctuates, leading to the fluctuation of the frequency  $\omega\big(I\big)$. Maximal values of the deviation of action was  evaluated in  \cite{UgMc11} and reads $\big(\delta I\big)_{max}\approx\sqrt{V \big(\frac{d\omega}{dI}\big)^{-1}}$ . Comparing maximal values of fluctuation $\big(\delta I\big)_{max}$ with the maximal values of the action increment observed for the dispersive map (22) we deduce $\frac{\big(\delta I\big)_{max}}{\sqrt{\big(\Delta I\big)^{2}}}\approx \frac{2\sqrt{\gamma}}{K^{3/2}}$. Typical values of the dimensionless dispassion constant is small $\gamma <<1$, while stochastic parameter is large
$K>1$. Since nonlinear frequency  is linear function of action for the frequency we have the same estimation $\frac{\big(\delta \omega\big)_{max}}{\sqrt{\big(\Delta \omega\big)^{2}}}\approx \frac{\big(\delta I\big)_{max}}{\sqrt{\big(\Delta I\big)^{2}}}= \frac{2\sqrt{\gamma}}{K^{3/2}}<<1$. Here $\big(\delta \omega\big)_{max}$ is the maximal fluctuation of the frequency due to the fluctuation of action, and  $\sqrt{\big(\Delta \omega\big)^{2}}$ is the maximal diffusive growth of the frequency. Consequently we may conclude that effect of fluctuations is reasonable small and accuracy of the frequency switching is high.

\section{Ensemble of oscillators}
\begin{figure}
    \centering
    \includegraphics[width=.74\textwidth]{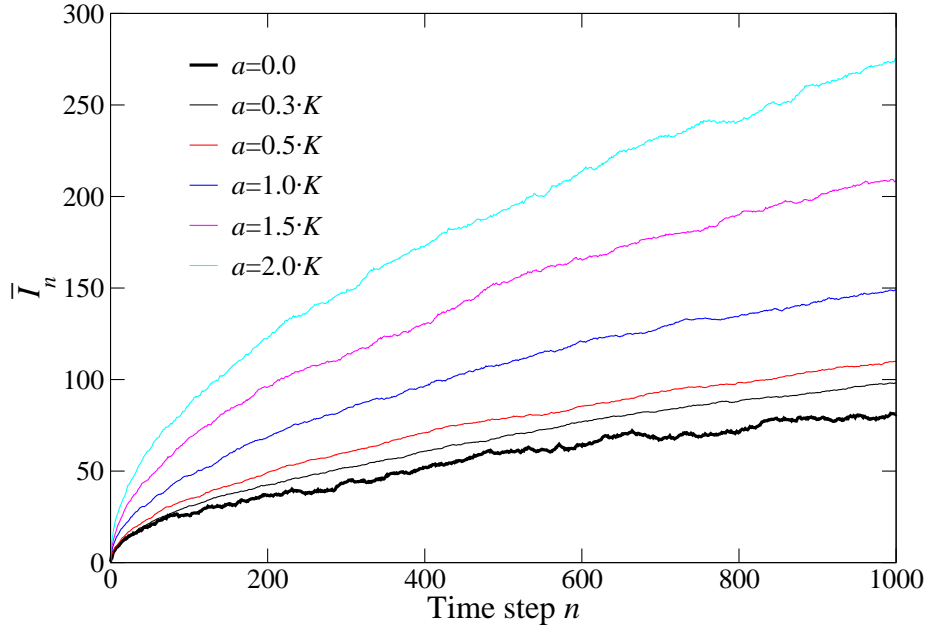}
    \caption{$\overline{I}_n$-dynamics of a large chain of $1000$ oscillators on the scale $n\in [0:1000]$ for $K=5.0$, $\gamma=0.0$, $\xi=0.0$ and various interaction strengths $a$. The initial configuration is randomly  chosen from the intervals $I_0^i
    \in (0:1)$ and $\theta_0^i \in (0:2\pi)$. $\overline{I}_n$ is defined according to Table \ref{tab_1}.}
    \label{fig_7}
    \end{figure}

\begin{figure}
    \centering
    \includegraphics[width=.74\textwidth]{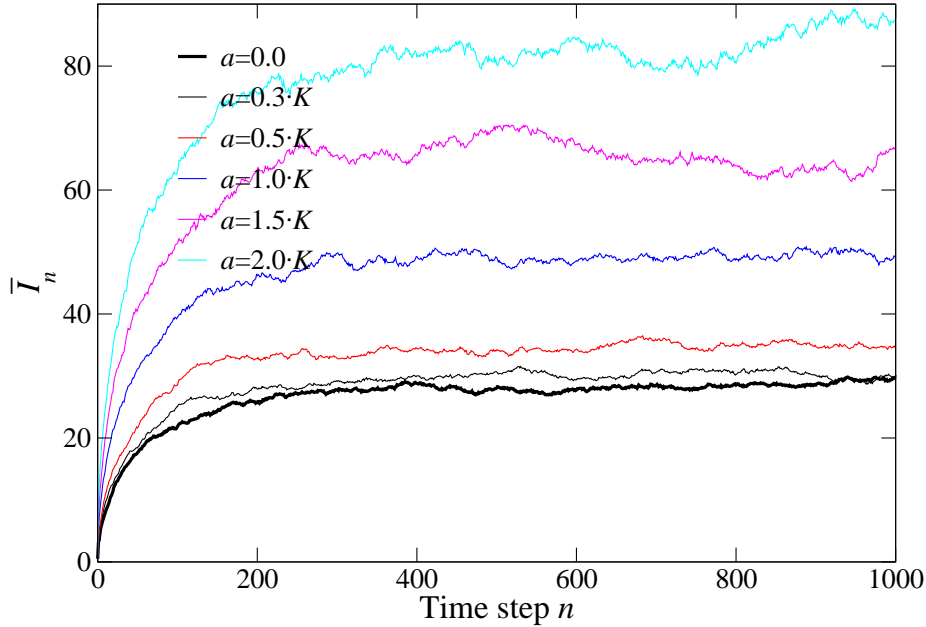}
    \caption{$\overline{I}_n$-dynamics of a chain of $1000$ oscillators on the scale $n\in [0:1000]$ for $K=5.0$, $\gamma=0.005$, $\xi=0.0$
      and various interaction strengths $a$. Initial configurations are as in Fig.~\ref{fig_7}.
      $\overline{I}_n$ is defined according to Table \ref{tab_1}.}
    \label{fig_8}
\end{figure}

\begin{figure}
    \centering
    \includegraphics[width=.74\textwidth]{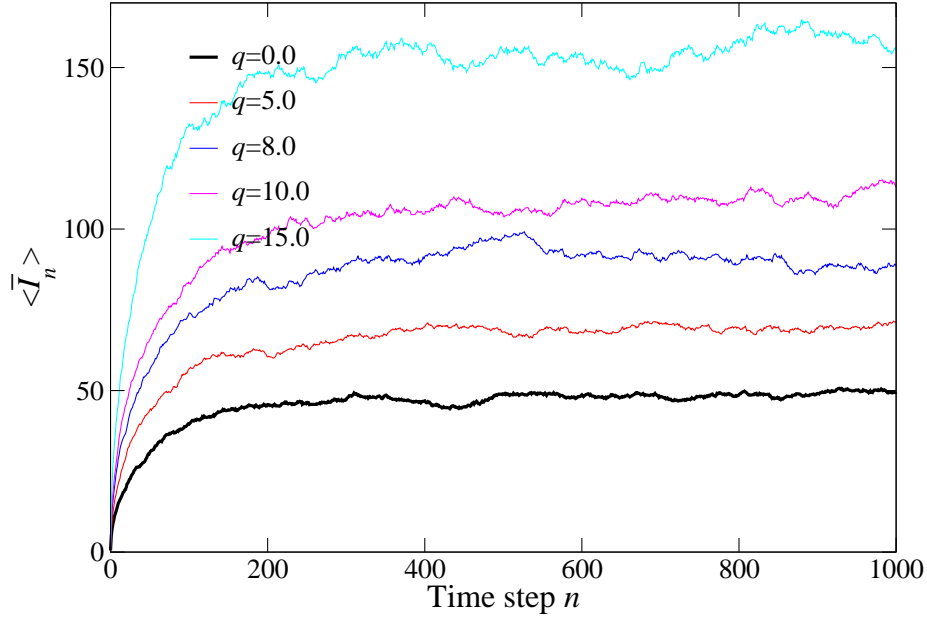}
    \caption{$<\overline{I}_n>$ dynamics for 1000 oscillators on the scale $n\in [0:1000]$ for fixed $K=5.0$,
    $\gamma=0.005$, and the interaction strength $a=1.0\cdot K$ for various noise levels $q$.
      Initial configurations as in Fig.\ref{fig_7}.}
    \label{fig_9}
\end{figure}

Here we generalize our approach to a chain of interacting oscillators. 
In non-integrable Hamiltonian systems with a large number of degrees of freedom $F$, the invariant KAM torus cannot isolate the stochastic layers formed in the vicinity of the separatrix lines \cite{WoLi90,KaKo89,GyLi89,ChTo09,MuAh11,LaTo99}.
Naively, one can assume that the large number of degrees of freedom leads to stronger chaos. In principal, such a statement is correct, however, even in the presence of strong chaos the phase space of the system still may contain  regular islands.  However, as demonstrated in Ref. \cite{GyLi89} the measure of the regular islands decays exponentially with the dimension of the system. 
For real materials, we assume that $F\gg 1$ and consider the effect of dissipation and noise as in the $F=1$ case in Sec. II. Keeping in mind the experimental situation, we will pay a special attention and precisely evaluate the time dependence of the diffusion coefficient.
We assume that the interaction between the oscillators is mediated by an external driving field and generalize expression (\ref{eq_5}) for the multidimensional case
\begin{equation}
\begin{array}{l}
\displaystyle H = \sum\limits_{i = 1}^N {\left( {H_0 (I_i ) + V(I_i ,\theta _i ,t)} \right)}, \\
\displaystyle  H_0 (I_i ) = I_i \omega _0  + 3\pi \beta \left( {\frac{{I_i }}{{m\omega _0 }}} \right)^2, \\
\displaystyle  V(I_i ,\theta _i ,t) = V(I_i ,\theta _i )T\sum\limits_{n =  - \infty }^\infty  {\delta (t - Tn)} , \\
\displaystyle   \mathrm{where}\,\,\, V(I_i ,\theta _i ) = V_0 (I_i )\cos \theta _i,\,\,\, V = V_0 (I_i ) \approx V_0 \sqrt {\frac{{I_i (0)}}{{m\omega _0 }}},\\
\displaystyle  \omega (I_i ) = \omega _0  + \frac{{6\pi \beta }}{{m^2 \omega _0^2 }}I_i .
\end{array}
\label{eq_34}
\end{equation}
Eqs. (\ref{eq_34}) generates the map that can be considered as a generalization of the K. Kaneko, and T. Konishi model \cite{KaKo89}
\begin{equation}
\begin{array}{l}
\displaystyle I_{n + 1}^i  = I_n^i  + K\sin \theta _n^i  - a \cdot \sin \left( {\theta _n^{i + 1}  - \theta _n^i } \right) + a \cdot \sin \left( {\theta _n^i  - \theta _n^{i - 1} } \right), \\
\displaystyle \theta _n^{i + 1}  = \omega _0 T + \theta _n^i  + I_{n + 1}^i. \\
\end{array}
\label{eq_35}
\end{equation}
In the linear limit when the deflection  of the angles from the equilibrium position $\theta_{n}^{i}$ is small and $\sin\theta_{n}^{i}$ can be linearized, the system (\ref{eq_35}) has a solution of the type of localized breathers \cite{AbFl98}. However,  we are interested in the strongly nonlinear chaotic case. The results of a numerical integration of the map given by eq. (\ref{eq_35}) are presented in Fig. \ref{fig_7}, which shows that, when increasing the coupling strength between the oscillators, the diffusion process is indeed enhanced.\\
Following the standard approach already used above and including dissipation we obtain the new map
\begin{equation}
\begin{array}{l}
\displaystyle I_{n + 1}^i  = e^{ - \gamma } I_n^i  + K(\gamma )\sin \theta _n^i  - a(\gamma ) \cdot \left(\sin ( {\theta _n^{i + 1}  - \theta _n^i }) - \sin ( {\theta _n^i  - \theta _n^{i - 1} }) \right), \\
\displaystyle  \theta _n^{i + 1}  = \theta _n^i  + I_{n + 1}^i, \\
\displaystyle  K(\gamma ) = \left( {\frac{{1 - e^{ - \gamma } }}{\gamma }} \right)K, \\
\displaystyle a(\gamma ) = \left( {\frac{{1 - e^{ - \gamma } }}{\gamma }} \right)a(\gamma ),
\end{array}
\label{eq_36}
\end{equation}
\textcolor{black}{where $a(\gamma)$ is the oscillators interaction strength.}\\
Results of numerical integrations of the dissipative map given by eq. (\ref{eq_36}) are presented in Fig. \ref{fig_8}. The time behavior of the diffusion process for the case of coupled oscillators is qualitatively the same as for the case of a single oscillator. Namely, the saturation of the diffusion is observed. The novelty is, however, that the saturated value of the action grows with increasing coupling between the oscillators.
\newline By analogy to the standard map of eq. (\ref{eq_31}), the thermal noise can be included as follows
\begin{equation}
\begin{array}{l}
\displaystyle I_{n + 1}^i  = e^{ - \gamma } I_n^i  + K(\gamma )\sin \theta _n^i  - a(\gamma ) \cdot \left( \sin ( {\theta _n^{i + 1}  - \theta _n^i } ) - \sin ( {\theta _n^i  - \theta _n^{i - 1} } ) \right), \\
\displaystyle  \theta _n^{i + 1}  = \theta _n^i  + I_{n + 1}^i, \\
\displaystyle  K(\gamma ) = \left( {\frac{{1 - e^{ - \gamma } }}{\gamma }} \right)\left( {K + \xi (t)} \right), \\
\displaystyle a(\gamma ) = \left( {\frac{{1 - e^{ - \gamma } }}{\gamma }} \right)\left( {a(\gamma ) + \xi (t)} \right).
\end{array}
\label{eq_37}
\end{equation}
Numerical results from Eqs.~(\ref{eq_37}) are presented in Fig. \ref{fig_9}.

We see that in the case of ensemble of oscillators  thermal noise again results in increased saturated values of action. However  absolute saturated values observed for the ensemble of oscillators are bit larger than values corresponding to the single oscillator case see Fig. \ref{fig_6}. Therefore
we can conclude that nonzero interaction strength $a(\gamma)$ (cf. Fig. \ref{fig_8}) enhances the diffusion process. \\

\begin{figure}
    \centering
    \includegraphics[width=.74\textwidth]{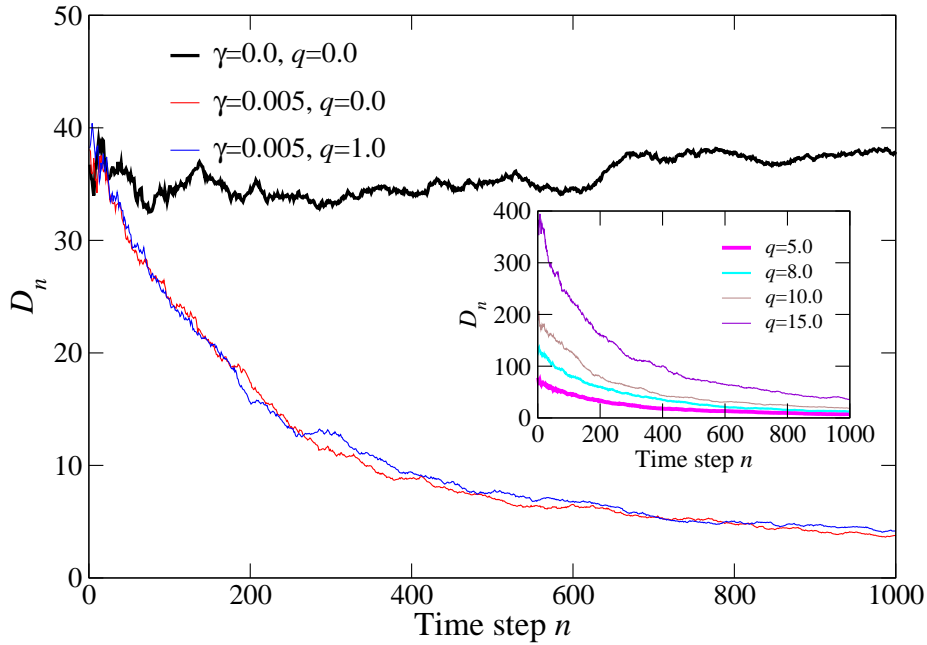}
    \caption{Time propagation of the diffusion coefficient $D_n$ given by eq. (\ref{eq_38}) in the non-dissipative and noise-free case (thick black curve), in the case of small dissipation $\gamma=0.005$ (thick red curve) and in the case where both dissipation $\gamma=0.005$ and thermal noise $q=1.0$ are present (thin blue curve). The inset demonstrates the effect of growing thermal noise.}
    \label{fig_10}
\end{figure}

We also numerically evaluate diffusion coefficient using the following expression
\begin{equation}
D_n \equiv D(t = nT) = \left\langle {\frac{1}{N}\sum\limits_{i = 1}^N {\frac{1}{n}\left[ {I_n^i  - I_0^i } \right]^2 } } \right\rangle .
\label{eq_38}
\end{equation}
When the diffusion in the phase space is normal, there exists a finite constant $D_\infty   = \lim _{t \to \infty } D(t)$. Otherwise, in the case of an anomalous diffusion we have a polynomial decay $D_\infty   = 1/t^\delta  ,\,\,\,\,\delta  > 1$ \cite{KaKo89}.\\
The results of the numerical integration for eq. (\ref{eq_38}) are presented in Fig. \ref{fig_10}. We infer from the figure that in the absence of the dissipation the diffusion is normal and the diffusion coefficient is a constant. The presence of the dissipation leads to a decay of the diffusion coefficient, confirming thus the saturation of the outgoing frequency. Hence, the considered chain of nonlinear oscillators allows for a controlled energy absorption, and may act as a prototype of a transistor in the optical frequency domain.

\section{Conclusions}
The main finding of the present study is that the maximum frequency of the optical transistor is given by expression (\ref{eq_32}), i.e.
$\displaystyle \omega _{\max } \left( {t \to \infty } \right) \approx \omega _0  + 24\pi \omega _0 \sqrt {1 + \frac{{\left( {K + \alpha_{0} q} \right)^2 }}{{4\gamma }}}$ where $\omega_0$ is the initial linear frequency, and the coefficient $K$ is the coefficient of the stochastisity, which can be easily controlled by the amplitude of the applied pulses $f$ and the intervals between them $T$. The coefficient $\gamma$ describes the phenomenological decay constant, $q$ defines the strength of the noise and $\alpha_{0}$ is a numerical factor which decays with increasing dissipation in the system  $\alpha_{0} (\gamma=0.001)\approx 19$, $\alpha_{0} (\gamma=0.005)\approx 9$ and $\alpha_{0} (\gamma=0.01)\approx 7$ (See Fig.6).
The observed switching time is equal to about $100 T$, i.e. $T\approx 10^{-12}$~[s]. The amplitude of the applied pulses should satisfy the condition $f>10^7$~[V/m], which is realistic for the $Nd$ glass lasers \cite{Yari76,Bloe65}.\\
In addition to a single oscillator model, we have also considered a model of coupled oscillators, where for both models the inclusion of noise into the applied electric pulses is shown to be an efficient tool for increasing the maximum frequency of the optical transistor.

\section{Acknowledgements}
The financial support by the Deutsche Forschungsgemeinschaft (DFG) through SFB 762, the grant No. SU 690/1-1, the HGSFP (grant No. GSC 129/1), and STCU grant No. 5053 is gratefully acknowledged.

\section*{Appendix}
\begin{table}[!h]
\caption{Definitions employed for calculations.}
\centering
\vspace{0.ex}
\begin{tabular}{c|c}
\hline
\hline
Single body problem $N=1$ \tabrule & Many body problem $i=1, ..., N$, $N=1000$\\ \hline \hline
$\bullet$ one realization: \tabrule & $\bullet$ one realization: \\
$\displaystyle I_{n}\equiv \sqrt{I^2_n}$ \tabrule & $\displaystyle \overline{I}_n\equiv \frac{1}{N}\sum_{i=1}^N\Big(I_n\Big)^i=\frac{1}{N}\sum_{i=1}^N\sqrt{\Big(I^i_n\Big)^2}$ \\
$\bullet$ $\xi\neq 0$, many realizations: \tabrule & $\bullet$ $\xi\neq 0$, many realizations: \\
$\displaystyle \sqrt{<I^2_n>}\equiv \sqrt{\frac{1}{R}\sum_{r=1}^R\Big(I^r_n\Big)^2}$, \tabrule & $\displaystyle <\overline{I}_n>\equiv \frac{1}{N}\sum_{i=1}^N\sqrt{\frac{1}{R}\sum_{r=1}^R\Big(I_n^{ri}\Big)^2}$, \\
$R=1000$ \tabrule & $R=1000$ \\
\hline \hline
\end{tabular}
\label{tab_1}
\end{table}

\end{document}